\documentclass[8pt]{article}
\usepackage[utf8]{inputenc}
\usepackage[affil-it]{authblk}
\usepackage[a4paper, total={7in, 10in}]{geometry}
\usepackage[symbol]{footmisc}
\usepackage{ragged2e}
\usepackage{titlesec}
\titleformat*{\subsection}{\normalfont\bfseries}
\titleformat*{\subsubsection}{\normalfont\itshape}
\usepackage{multicol}
\usepackage[format=plain,font=small,labelfont=bf,textfont=bf]{caption}

\usepackage{framed,multirow}
\usepackage{amssymb}
\usepackage{latexsym}
\usepackage{url}
\usepackage{xcolor}
\usepackage{hyperref}
\usepackage{amsmath,amssymb,amsfonts}
\usepackage{algorithm}
\usepackage{algpseudocode}
\usepackage{graphicx}
\usepackage{textcomp}
\usepackage{multirow}
\usepackage{booktabs}
\usepackage{tabu}
\usepackage{tabularx}
\usepackage{dcolumn} 
  \newcolumntype{d}[1]{D{.}{.}{#1}}    
  
\makeatletter
\newenvironment{tablehere}
  {\def\@captype{table}}
  {}

\newenvironment{figurehere}
  {\def\@captype{figure}}
  {}
\makeatother

\makeatletter
\renewcommand{\maketitle}{\bgroup\setlength{\parindent}{0pt}
\begin{flushleft}
  \textbf{\@title} \vspace{2em}

  \@author
  
\end{flushleft}\egroup
    
}
\makeatother

\title{\huge{Evaluation of automated airway morphological quantification for assessing fibrosing lung disease}}

\author[1,2]{Ashkan Pakzad\footnote{Corresponding author: 
email: a.pakzad@cs.ucl.ac.uk}}
\author[1,3]{Wing Keung Cheung}
\author[1,2]{Kin Quan}
\author[4]{Nesrin Mogulkoc} 
\author[5]{Coline H.M. Van Moorsel} 
\author[6]{Brian J. Bartholmai} 
\author[7]{Hendrik W. Van Es} 
\author[4]{Alper Ezircan} 
\author[5]{Frouke Van Beek} 
\author[5,8]{Marcel Veltkamp} 
\author[9]{Ronald Karwoski} 
\author[10]{Tobias Peikert} 
\author[10]{Ryan D. Clay}
\author[10]{Finbar Foley} 
\author[10]{Cassandra Braun}
\author[11]{Recep Savas}
\author[1,12]{Carole Sudre}
\author[13]{Tom Doel}
\author[1,3]{Daniel C. Alexander}
\author[1,3]{Peter Wijeratne} 
\author[1,2]{David Hawkes} 
\author[1,2]{Yipeng Hu} 
\author[14]{John R Hurst} 
\author[1,14]{Joseph Jacob}

\affil[1]{Centre for Medical Image Computing, University College London, UK} 
\affil[2]{Department of Medical Physics and Biomedical Engineering, University College London, UK} 
\affil[3]{Department of Computer Science, University College London, UK} 
\affil[4]{Department of Respiratory Medicine, Ege University Hospital, Izmir, Turkey} 
\affil[5]{Department of Pulmonology, Interstitial Lung Diseases Center of Excellence, St Antonius Hospital, Nieuwegein, the Netherlands} 
\affil[6]{Division of Radiology, Mayo Clinic Rochester, Rochester, MN, USA} 
\affil[7]{Department of Radiology, St Antonius Hospital, Nieuwegein, the Netherlands} 
\affil[8]{Division of Heart and Lungs, University Medical Center, Utrecht, the Netherlands} 
\affil[9]{Department of Physiology and Biomedical Engineering, Mayo Clinic Rochester, Rochester, MN, USA} 
\affil[10]{Division of Pulmonary and Critical Care Medicine, Mayo Clinic, Rochester, MN, USA} 
\affil[11]{Department of Radiology, Ege University Faculty of Medicine,  Izmir, 35100, Turkey} 
\affil[12]{MRC Unit for Lifelong Health \& Ageing, University College London, UK} 
\affil[13]{Code Choreography Limited, UK} 
\affil[14]{UCL Respiratory, University College London, UK} 

\begin{document}
\begin{titlepage}
\maketitle
\centering \noindent\rule{8cm}{0.4pt}
\section*{\centering Abstract}
\justify 
\large{
Abnormal airway dilatation, termed traction bronchiectasis, is a typical feature of idiopathic pulmonary fibrosis (IPF). Volumetric computed tomography (CT) imaging captures the loss of normal airway tapering in IPF. We postulated that automated quantification of airway abnormalities could provide estimates of IPF disease extent and severity. 

We propose AirQuant, an automated computational pipeline that systematically parcellates the airway tree into its lobes and generational branches from a deep learning based airway segmentation, deriving airway structural measures from chest CT. Importantly, AirQuant prevents the occurrence of spurious airway branches by thick wave propagation and removes loops in the airway-tree by graph search, overcoming limitations of existing airway skeletonisation algorithms. Tapering between airway segments (intertapering) and airway tortuosity computed by AirQuant were compared between 14 healthy participants and 14 IPF patients. 

Airway intertapering was significantly reduced in IPF patients, and airway tortuosity was significantly increased when compared to healthy controls. Differences were most marked in the lower lobes, conforming to the typical distribution of IPF-related damage. 

AirQuant is an open-source pipeline that avoids limitations of existing airway quantification algorithms and has clinical interpretability. Automated airway measurements may have potential as novel imaging biomarkers of IPF severity and disease extent. 
}
\end{titlepage}
\newgeometry{a4paper, total={7.5in, 10in}}

\begin{multicols}{2}

\section{Introduction}

We present a clinical tool for the comprehensive evaluation of airway structure on volumetric Computed Tomography (CT) imaging of the lungs. We apply this tool to quantify traction bronchiectasis in idiopathic pulmonary fibrosis. Taking as input the CT image and a detailed airway segmentation, the tool outputs quantitative metrics on each airway segment, indexed by lung lobe and airway generation.

\subsection{Bronchiectasis}

Airways are tubular branching structures originating centrally from the trachea and extending to the lung periphery. Airways transport gases between the external air and the alveolar sacs at the airway terminus where gas exchange into the alveolar capillaries occurs.
An airway segment is defined as a continuous tube running between two airway branching points. From the major bronchi that arise from the trachea, each new division of airway branches can be considered a new airway generation. In a healthy individual, airway segments narrow or taper in diameter as they run from the central to the peripheral lung. Tapering occurs both along an airway segment and with respect to the segment of the preceding airway generation \cite{weibel_architecture_1962}.

Bronchiectasis describes a structural airway disease in which the airways lose their healthy tapered structure and become abnormally dilated within a segment. Various lung diseases are associated with bronchiectasis, including those driven by infection and or inflammation in the airway wall. In fibrosing lung diseases, of which idiopathic pulmonary fibrosis (IPF) is the hallmark fibrosing lung disease \cite{hansell_imaging_2009} the airways are pulled open by fibrosis and contraction of the surrounding connective tissue and airway dilatation is termed 'traction bronchiectasis'.

Bronchiectasis is typically evaluated by a  radiologist following visual inspection of a chest CT scan. Evaluation of the extent and degree of dilatation of bronchiectatic airways allows the characterisation of lung disease severity and extent. In IPF for example, the prognostic importance of airway dilatation has influenced current diagnostic guidelines with the presence of traction bronchiectasis in an appropriate distribution now being classified as a probable usual interstitial pneumonia pattern  \cite{raghu_diagnosis_2018}. 

The classical morphological signs of bronchiectasis include a) the visualisation of airways within 1 cm of the lung periphery, b) a lack of tapering of the airway, c) an airway diameter greater than the diameter of the accompanying pulmonary artery \cite{hansell_fleischner_2008}. 
However, solely relying on the comparison of the airway to its adjacent pulmonary artery can be misleading. Living at high altitude \cite{kim_bronchoarterial_1997} and normal ageing \cite{matsuoka_bronchoarterial_2003} can result in non-pathological airway dilation which can be confused with bronchiectasis. Furthermore, there are pathological mechanisms that can result in changes to the size of the pulmonary artery such as smoking \cite{diaz_quantitative_2017} and hypoxia-induced pulmonary vasoconstriction as a result of chronic lung disease \cite{dunham-snary_hypoxic_2017}.
Typically the visual markers of bronchiectasis are assessed for severity on an ordinal scale and on a lobar basis \cite{bhalla_cystic_1991}. However, these scoring systems lack sensitivity, are time consuming to apply and are associated with disagreement between radiologists. Consequently they are not used in routine clinical practice or research.
Moreover, the ordinal scores conflate disease extent (number of involved lobar segments) and severity (degree of abnormal airway dilatation) which could potentially dilute the prognostic signal attributable to disease extent or severity individually.

Computational image analysis of lung CT imaging may allow the derivation of objective robust quantitative measures of abnormal airway dilatation extent and severity by quantifying dilatation to the nearest millimetre. The precise quantification of airway damage, may identify IPF patients at risk of rapid disease progression. Identification of such patients would be an important cohort enrichment strategy for recruitment into clinical trials of novel IPF therapies \cite{johannson_models_2015,collard_new_2015}.

\subsection{Idiopathic Pulmonary Fibrosis}
IPF is a lung disease characterised by excessive fibrosis in the structural framework of the lung. The fibrotic process characteristically causes traction bronchiectasis in the lung periphery. There are 80,000 new cases of IPF diagnosed every year across Europe and the United States \cite{hutchinson_global_2015} with patients typically surviving only 3-5 years following diagnosis \cite{spencer_idiopathic_2021}.

\subsection{Previous Works}
Acquiring tapering measurements involves executing an airway measurement algorithm at a perpendicular plane to the airway centreline at regular intervals, this is demonstrated in figure \ref{fig:tapering}(a). To date, two methods have been used to measure the tapering gradient of airways on CT imaging. Both methods require an airway lumen segmentation and airway centreline. The first utilises the full width at half maximum edgecued segmentation limited (FWHM\textsubscript{ESL}) technique, where a one dimensional profile through the airway wall from inside the airway lumen to outside the airway wall would trace a bell shaped curve. The boundary edges lie at the half-maximum value \cite{kiraly_virtual_2005, quan_tapering_2018, odry_automated_2006}. The second method of measuring an airway tapering gradient aims to optimise an initial airway lumen segmentation to better fit the luminal airway surface on the image and the surface of the outer airway wall \cite{petersen_optimal_2014}.

The methodology used in this manuscript utilises and expands on methods laid out in \cite{quan_tapering_2018}. The paper by Quan et al demonstrated a pipeline that calculated the airway tapering gradient from the carina (where the major bronchi split) to the distal-most point of an airway branch segmentation. The pipeline was validated following experiments demonstrating accurate tapering measurements on bespoke CT phantom tubes which encompassed airway tapering gradients in the range of $4\% $ to $16\%$ down to $2.5mm$ diameter \cite{quan_tapering_2018}. 

The method by \cite{kuo_airway_2020} was shown to be able to identify differences in airway tapering in a small cohort of paediatric cystic fibrosis patients. The methodology allowed for global airway analysis, using the intertapering measurement. However, for successful execution of the method, time-consuming manual airway segmentations and skeletonisations were required. 

\begin{figurehere}
    \centering
    \includegraphics[width=\columnwidth]{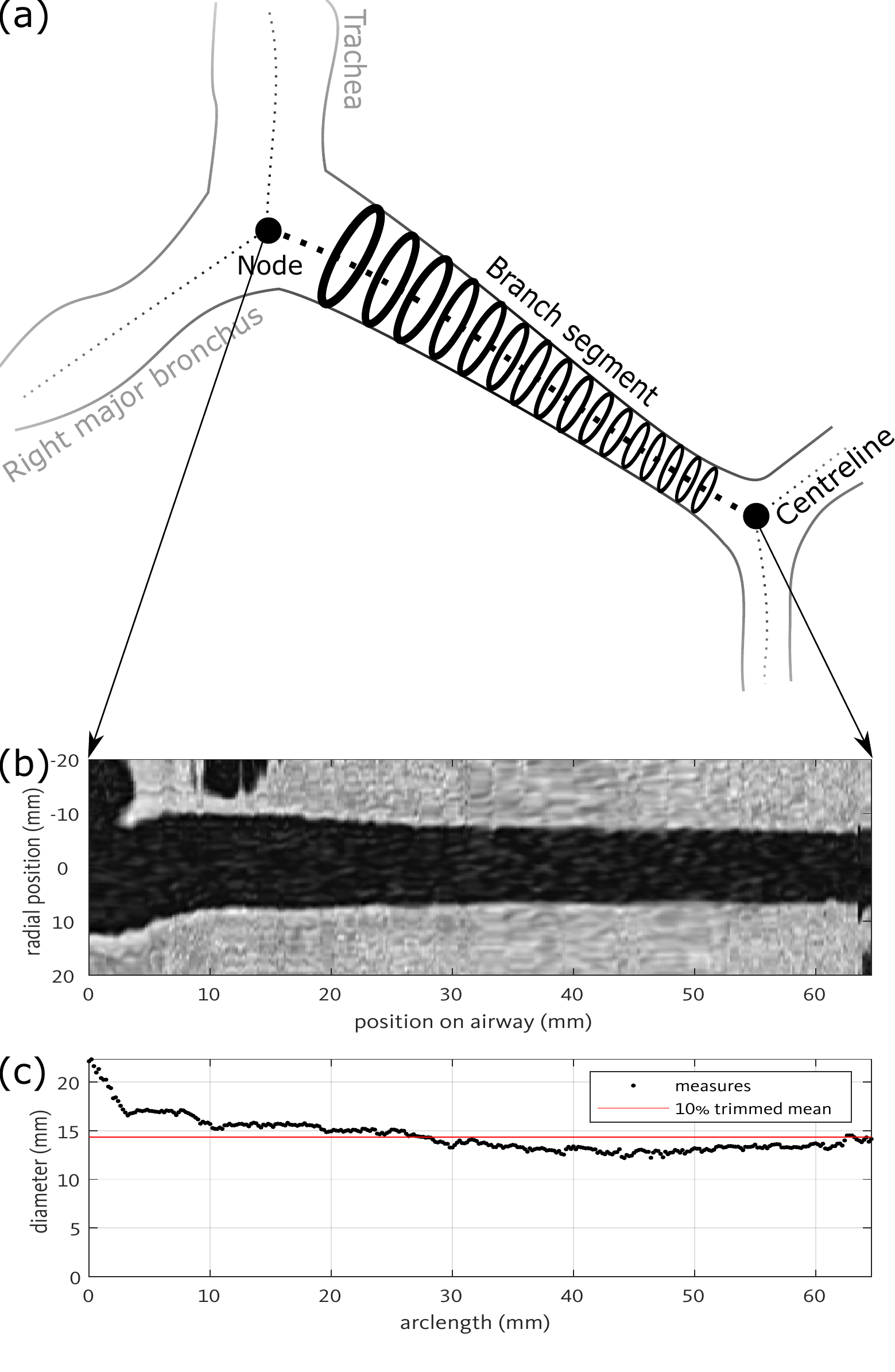}
    \caption{Example of how tapering measures are derived for a given airway, here the left major bronchus is demonstrated. (a) The airway is outlined with a solid line with its segmentation-derived centreline represented by a dashed line. An airway segment is typically bounded by nodes which represent splitting of airways or end points. The airway diameter is measured at regular intervals between nodes along the segment, shown here as bold rings.
    (b) Our technique uses the reformatted CT interpolated along the airway segment's centreline between nodes. (c) shows corresponding diameter measured using the presented pipeline, AirQuant. The ten percent trimmed-mean, i.e. uses the central 90\% of diameter measures to derive the mean, highlighted in red.}
    \label{fig:tapering}
\end{figurehere}

\vspace{2em}

\subsection{Proposed airway measurement pipeline}
In this paper, we adapt existing segmentation and skeletonisation methods to reliably extract airway centrelines and branching tree-like structures. To replicate lung lobar classification systems used in standard radiological assessment, we have developed an automated lobar classification system that delineates the lingula (nominally part of the left upper lobe) as a distinct lobe.

Our methodology results in the first automated pipeline capable of measuring inter-branch tapering and tortuosity of all segmented airways on a lung CT. Inter-branch tapering, known as intertapering is the relative percentage change in average diameter of an airway segment compared to its parent segment. This measure was first considered by \cite{kuo_airway_2020}, though their method relied on manually drawn centrelines, with a heavy labour cost on the user, when deriving tapering measures.

We propose an end-to-end pipeline with the motivation of providing an objectively derived measure of airway morphology. It requires little to no manual input, making it a feasible clinical tool to measure airways in order to support clinical assessment. 

The lungs are classified by lobe and by airway generation allowing easy localisation of focal airway abnormalities. We describe various quantitative expressions of airway morphological abnormality in IPF patients that delineate the extent and severity of fibrosis-related airway damage. 

We also show that a deep learning regression approach to calculate airway inter-branch tapering and tortuosity directly from CT images failed to generalise, thereby validating our end-to-end pipeline approach. The comparison evinced an application where modern end-to-end deep neural networks, that are perceived to be more powerful, could not and in all likelihood should not replace the proposed hand-engineered solution. Our hand-engineered solution generalised more robustly, and demonstrated superior clinical performance on unseen data.

We contrast findings in IPF with airway metrics in healthy volunteers. Notably, we measure and report airway tortuosity for the first time.

\section{Methods}
\begin{figure*}[!t]
    \centering
    \includegraphics[width=\textwidth]{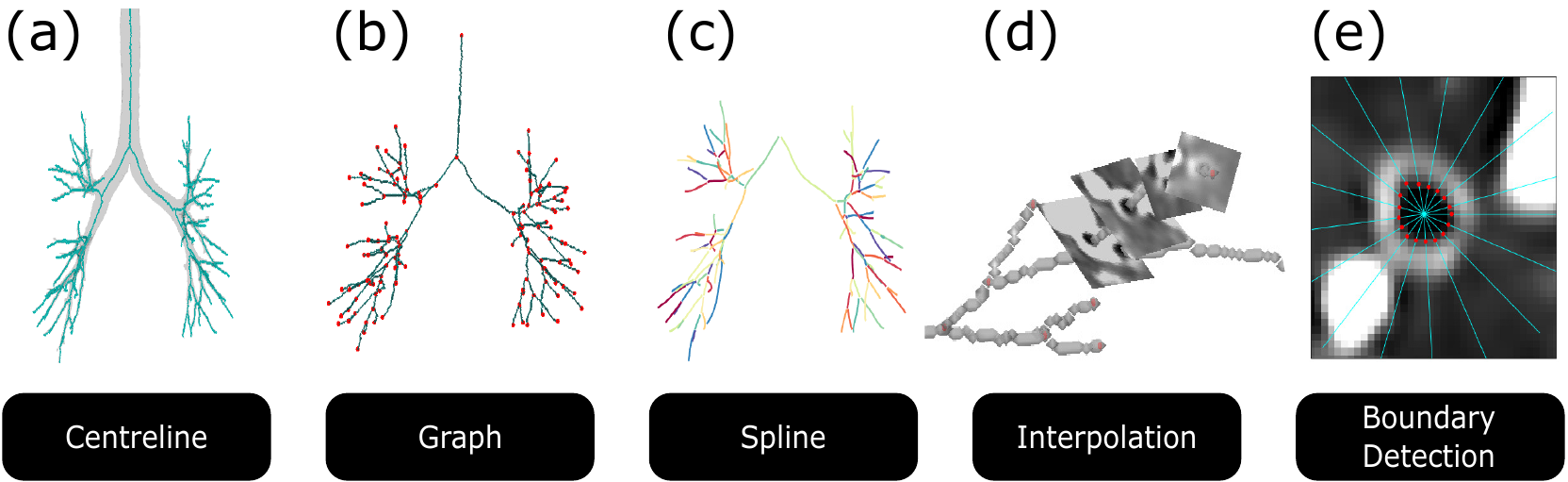}
    \caption{Demonstrating the AirQuant pipeline graphically from left to right for a given airway lumen segmentation through to the end where the lumen boundary is established.}
    \label{fig:pipeline}
\end{figure*}

Following automated airway segmentation based on a trained dilated 2D-UNet combined with region-growing, the proposed overall pipeline to process a chest CT is shown in figure \ref{fig:pipeline}. The centreline is first extracted by thinning (figure \ref{fig:pipeline}(a)) and a network graph is derived. The airway tree is parcellated into its individual branching segments where graph edges and splitting/end points are represented by nodes. (figure \ref{fig:pipeline}(b)). Cubic splines are then fitted to each segment's associated centreline points, allowing sub-voxel interpolation along the airway segment ((figure \ref{fig:pipeline}(c)). By taking the tangent to the spline at intervals, CT patches in the plane perpendicular to the airway centreline can be cubically interpolated (figure \ref{fig:pipeline}(d)). From these patches, the airway luminal diameter is measured using the (FWHM\textsubscript{ESL}) technique by \cite{kiraly_virtual_2005} (figure \ref{fig:pipeline}(e)). Using the set of luminal diameter measurements and total arc-length by spline fitting, metrics can be derived that describe each individual airway segment.

All code was written and executed in MATLAB \footnote{\url{https://github.com/ashkanpakzad/AirQuant}}.

\subsection{Study Data}
\label{sec:data}
We consider 14 healthy never-smokers (from Mayo Clinic, Rochester, Minnesota, USA) and 14 diverse IPF patients from two centres (11 from St Antonius Hospital, Utrecht, Netherlands and 3 from Ege University Hospital, Izmir, Turkey). Further technical details of the study data are included in table \ref{tab:data}. None of the cases analysed in this study were used to train or test the airway segmentation algorithm mentioned in section \ref{sec:2dunet}.
\begin{table*}[!b]
\centering
\caption{Technical details of the CT data used in this study. Values shown as median (inter quartile range) across cases where applicable. Significance shown for comparison of number of lobe segments by Mann-Whitney U test * $p < 0.05$ and *** $p < 0.001$.}
\vspace{1em}
\label{tab:data}
\begin{tabular}{l|l|l}
Feature              & Normal                                                              & Idiopathic pulmonary fibrosis                                                                     \\ \hline
         \hline
Number of cases                  & 14                                                                  & 14                                                                            \\
Slice pixel size       & 0.74 (0.70 - 0.74)                                                  & 0.72 (0.68 - 0.73)                                                            \\
Slice   thickness        & 1.00 (1.00 - 1.00)                                                  & 0.75 (0.70- 0.80)                                                             \\
Reconstruction   kernels & Bv49d                                                               & \begin{tabular}[c]{@{}c@{}}C, YC, BONE, EC, \\ YB,   L, STANDARD\end{tabular} \\
Total airway   segments  & \begin{tabular}[c]{@{}c@{}}128.50 (117.50 - 148.75)\end{tabular} & 160.50 (139.50 -   191.25)                                                    \\
\quad Right upper lobe         & 28.00   (26.25 - 30.00)                                             & 31.00 (27.25 -38.50)                                                          \\
\quad Left upper  lobe         & 20.00 (17.00 - 21.75)                                               & 21.50   (13.00 - 25.00)                                                       \\
\quad Right middle  lobe***    & 9.50 (7.00 - 11.75)                                                 & 15.50   (13.00 - 24.00)                                                       \\
\quad Left middle  lobe        & 7.00   (5.00 - 8.00)                                                & 7.50   (7.00 - 12.00)                                                         \\
\quad Right lower  lobe*       & 34.50   (30.25 - 38.00)                                             & 47.50 (42.50 - 53.00)                                                         \\
\quad Left lower lobe          & 32.00 (24.00 - 42.00)                                               & 40.50 (31.75 - 47.50)                                                        
\\
\bottomrule
\end{tabular}

\end{table*}

\subsection{Automated Airway Segmentation}
As an airway segmentation is required to measure airway tapering, we implement a trained 2D dilated UNet model \cite{yu_multi-scale_2016} and a region growing algorithm with explosion control using the software tool, Pulmonary ToolKit (PTK) \footnote{\url{https://github.com/tomdoel/pulmonarytoolkit}}. The two methods are executed on each case in parallel, with the results combined. Any objects unconnected to the airway tree are discarded.

\subsubsection{2D UNet Segmentation}
\label{sec:2dunet}
The dilated U-Net model is an improved version of the original U-Net model that replaces standard convolution layers with dilated convolution layers \cite{yu_multi-scale_2016}. The addition of dilated convolution layers allows the performance of local convolutional operations on a larger region without an increased computational cost but importantly maintains image resolution. Using this methodology provides greater pixel-wise context during training and at inference. Our model was trained on manually segmented airway trees performed in-house under the supervision of an experienced chest radiologist (J. Jacob). The training and validation/testing dataset used in the development of the dilated UNET model comprised six normal CTs in healthy never-smoker volunteers, two normal cases from the EXACT09 competition data set \cite{lo_extraction_2012} and 17 IPF cases, totalling 25 volumetric CTs. The axial slices of all CTs were amalgamated, randomised and split 80-20 into training and validation datasets, respectively. None of the images used to train/validate/test the dilated U-NET model were analysed in the current clinical study. As a 2D input model, the dilated U-NET only considered one axial CT slice at a time, without the context of the rest of the CT volume. Training was implemented using Adam optimiser \cite{kingma_adam_2017}, minimising the combined binary cross entropy \cite{goodfellow_deep_2016} and dice similarity coefficient loss functions \cite{dice_measures_1945}. The learning rate was initially set to 1e-5, reducing to 1e-6 upon loss plateau over three consecutive training epochs. The model achieved a training and validation accuracy of $88.5\%$ and $87.2\%$, respectively. Implementation was in Tensorflow \cite{martin_abadi_tensorflow_2015} and Keras \cite{chollet_keras_2015} with Python.

Data were preprocessed by limiting intensity levels to a standard lung window, level -500 Hounsfield Units (HU) and width 1500 HU, and then normalising to a range from 0 to 1. The slice size was limited to $512\times512$ pixels. Larger raw CT slices were downsampled by cubic interpolation to pass through the fixed-size model and inferred labels upsampled to the original size by nearest-neighbour interpolation. All analyses were conducted on the original image size.

\subsubsection{Pulmonary Toolkit Segmentation}
\label{sec:PTKseg}

The Pulmonary Toolkit (PTK) \cite{burrowes_combined_2017} implements airway segmentation by region growing from a tracheal seed. The algorithm starts by thresholding the CT to air voxel density. A wave-front propagates from the tracheal seed and travels through the thresholded mask, classifying the volume traversed as being part of the airway tree. The wave-front maintains a thickness of multiple voxels in the larger airways, which helps prevent spurious splitting of the wave-front due to noisy interior voxels. Complete splitting of a wave-front indicates new branching. Sudden increases in volume of wave growth indicates parenchymal leakage of the airway segmentation. Leakage is controlled by an \textit{explosion multiplier}, set such that if the number of new voxels in a wave front are more than the factor of the explosion multiplier of the previous wavefront, the algorithm defaults to its previous iteration. The default PTK settings used to avoid parenchymal leakage were implemented i.e. maximum number of generations was set to $15$ and the explosion multiplier to $7$.

\subsection{Centreline Extraction}
PTK's full skeletonisation method \cite{doel_developing_2012} is used here on the final airway segmentation. This is a thinning algorithm based on \cite{palagyi_quantitative_2006}. It first identifies airway endpoints by re-running the same wave propagation step as the airway segmentation algorithm described above \ref{sec:PTKseg}, without explosion control or generational limit. The thick wave-propagating component helps to delineate false airway branches from real branches, thereby identifying true airway endpoints. The trachea is marked as an airway endpoint from the outset. The binary object bounded by all endpoints is then iteratively reduced to its topological centreline. Although false branches are robustly removed, airway loops may have occurred at this stage, as topological thinning does not reflect the true anatomical tree structure of the airways. 

A key innovation of PTK's skeletonisation method is a post-processing step which removes inner loops by retracing out the whole skeleton voxel-by-voxel, parsing branch segments in a depth-first-search style. Starting from the top of the trachea, it considers the 27-connected neighbour skeleton voxels of the last traced voxel to determine which voxel should be added to grow the airway segment. Upon reaching the end of a segment, i.e. a bifurcation point, it has fully parsed the current segment and starts parsing one of two (or more) new child-branches. In doing so, it acknowledges the discovery, though does not immediately parse that child-branch's sibling(s). It holds in memory the first voxel of every new segment it discovers. If a candidate skeleton voxel is found to match the first voxel of a discovered non-parsed segment, then it identifies that the whole candidate segment forms a loop. The offending segment is then terminated and removed from the final skeleton.

\subsection{Airway Parcellation} 
\label{sec:parcellation}

\begin{figure*}[!t]
    \centering
    \includegraphics[width=\textwidth]{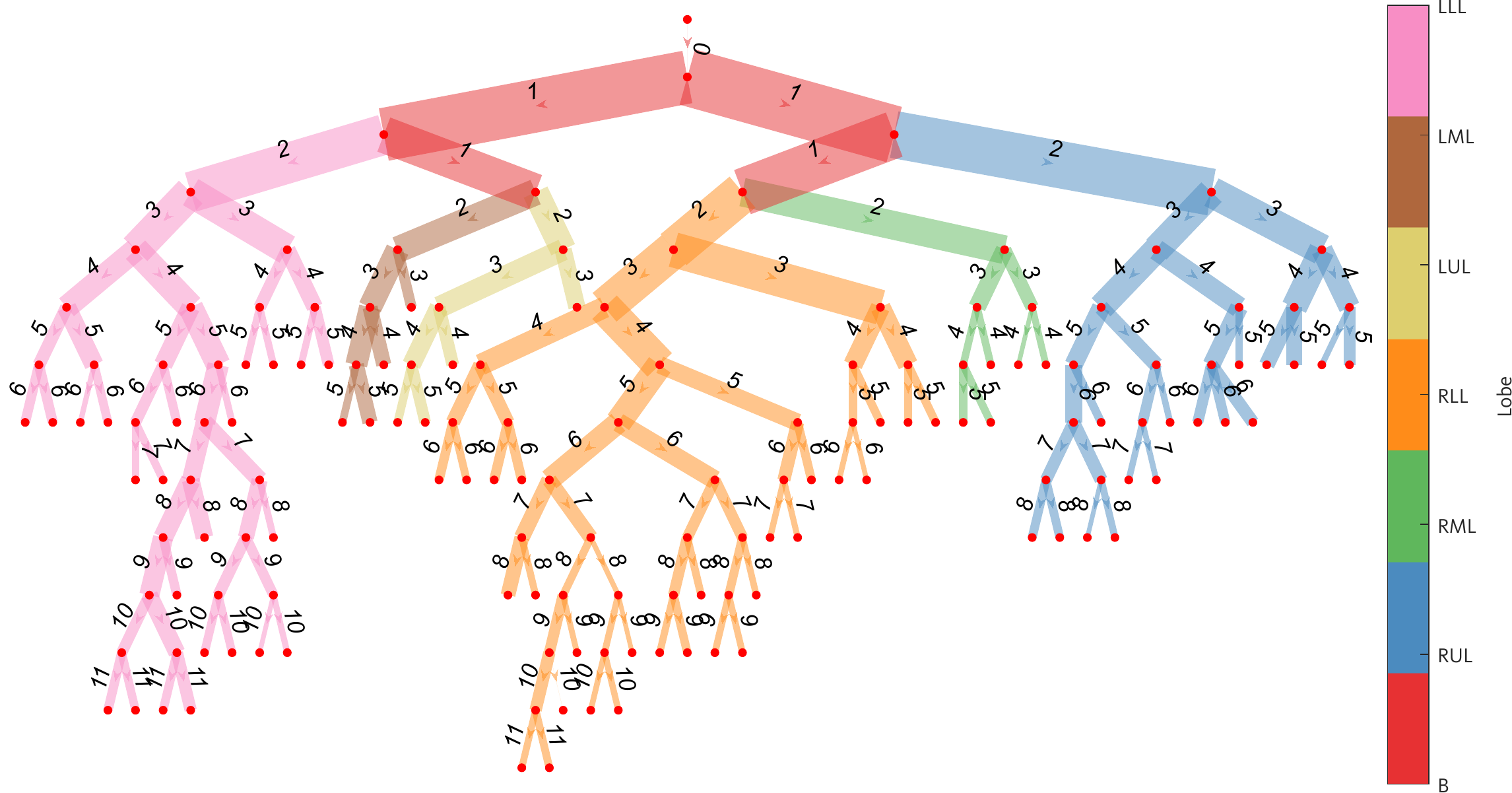}
    \caption{Example airway graph representation of an IPF patient with traction bronchiectasis, derived from volumetric computed tomography imaging using the pipeline presented in this manuscript. Airway divisions are represented by nodes and airway segments by edges. Edge thickness is proportional to average luminal diameter. The edge label represents the lobar generation of each airway segment: 0 for the trachea; 1 for the main and intermediate bronchi (B); 2 for the first intralobar airway; 3 onwards for each subsequent airway division. Edge colour is coded to lobe classification, RUL=right upper lobe, LUL=left upper lobe, RML=right middle lobe, LML=left middle lobe, RLL=right lower lobe, LLL=left lower lobe.}
    \label{fig:graph}
\end{figure*}

To facilitate analysis the airway tree is divided into airway segments. This definition enables direct conversion of the airway centreline into a graphical network representation \cite{kollmannsberger_small_2017}, where graph edges become airway segments and airway division points or airway endpoints become graph nodes (Figure \ref{fig:pipeline}(b)). The carina (point of division of the main bronchi from the trachea) can be identified by graph centrality. Conversion to a directional graph, setting the trachea as the origin with airways directed in the graph means that airway segments are directed from the central lungs outwards. Setting a direction in the graph facilitates subsequent steps in the pipeline, dictates the direction in which the spline is sampled and supports airway generation classification and lung lobe classification. Figure \ref{fig:graph} demonstrates the capabilities of this graph representation.

Each individual airway segment has a cubic spline (piecewise polynomial function) fitted to its collective centreline points which is smoothed by a moving average along the segment starting from the proximal segment and moving distally. The spline is sampled at equidistant intervals, tracking the tangent of the spline at these points to calculate the airway's perpendicular plane for diameter measurements.  The limit of resolution for the change in airway diameter is considered to be no less than half the shortest voxel diameter. All interpolation sampling sizes are set individually for each given CT image at half the shortest voxel diameter.


\subsection{Airway Measurement}
\label{sec:awy_measure}
The CT image is interpolated at spline sample points perpendicular to the tangent of the spline such that the resultant image produces a slice along the natural long axis of the airway (Figure \ref{fig:pipeline}(d)). The interpolation pixel size is dynamically set to half the shortest voxel diameter of the given CT. Diameter measurements are made on these airway-perpendicular slices at spline sampling intervals. 

On the airway-perpendicular slices, several radial density profiles that are uniformly spaced originate from the lumen centre, sampling the change in HU; a technique known as \emph{raycasting}. These radial profiles of the airway wall typically appear as a Gaussian curve, due to the nature of CT imaging of thin structures \cite{weinheimer_about_2008}. It is approximated that the wall centre falls at the Gaussian maximum and that the inner and outer boundary fall at the FWHM points, i.e. the half-intensity points either side of the curve. However, due to the nature of CT imaging and proximity of lung vessels to the airways, the radial profile does not always appear as a smooth Gaussian. In reality, these HU density profiles can appear noisy with several maxima. To improve robustness, the airway segmentation is also interpolated at the same points as the CT, the nearest local maxima to the boundary of the airway segmentation is considered the wall peak. This method is known as the FWHM\textsubscript{ESL} technique as described by \cite{kiraly_virtual_2005}, and implemented and validated by \cite{quan_tapering_2018} on phantoms down to $2.5mm$. The inner airway lumen boundary is therefore identified as the first half-maximum. The second half-maximum relating to the outer wall boundary is often found to be more susceptible to noise, as the outer airway wall may have a structure with similar contrast and density obscuring its boundary, e.g. a blood vessel. Thus we only consider the accurately determined inner boundary (Figure \ref{fig:pipeline}(e)). 

As a perpendicular section through an airway forms an ellipse, an ellipse is fitted \cite{fitzgibbon_direct_1996} to the inner wall boundary points. The ellipse total area, $A_{E}$ is calculated and a generalised diameter $D_{G}$ derived for that point in the airway segment.

\begin{align*}
D_{G} &= 2 \sqrt{\frac{A_{E}}{\pi}}
\end{align*}

Where our methodology differs from \cite{quan_tapering_2018} is that we employ an outlier detection mechanism improving robustness, demonstrated in figure \ref{fig:outlier}. The distance of each inner lumen point to the centre point is calculated, if this spans greater or less than 3 times the Mean Absolute Deviation (MAD) compared to all other points then it is considered an outlier. To support this, we choose to increase the number of raycast radial density profiles from 50 (every 7.2 degrees of rotation) to 180 (every 2 degrees of rotation). The greater the number of raycasts, the more robust outlier detection becomes, particularly for larger airways where the distance between lumen boundary points will be larger for a given number of raycasts. 

Ultimately, an accurate intertapering value relies on accurate average measurements of the current and parent airway segment. By sampling the airway diameter at less than half the smallest CT voxel, we are ensuring that we get the most accurate diameter measurements these methods can obtain from the CT image. Figure \ref{fig:tapering}(b,c) demonstrates a series of diameter measurements on a single airway segment.

\subsection{Airway Lobe classification}

Airway lobe classification is based on the method described and successfully tested on 300 COPD patients by \cite{gu_automated_2012}. 
The classification algorithm takes the graphical representation of the airway tree described in section \ref{sec:parcellation}. As decribed in algorithm \ref{alg:lobe1}, it considers the positional coordinates of nodes (see Figure \ref{fig:lobeclassgraph}) starting from the carinal node to identify the main left and right bronchi as well as the left upper (LUL) and lower lobes (LLL). The lingula lobe (left middle lobe (LML)) was identified as the lower branch in the left upper lobe thereby mimicking the lobar classification used in clinical radiological lung assessment. Edges (airway segments themselves) are immediately classified as off-spring of classified nodes.

\vspace{2em}

The right upper lobe (RUL) is the first division of the right main bronchus. The processing of classifying the right middle lobe (RML) and right lower lobe (RLL) is described in algorithm \ref{alg:lobe2}. The remaining non-classified right lung is taken as a subset of the overall airway graph and the difference in axial to lateral position of every end node is calculated. The most extreme nodes are considered to originate from the RML and the least extreme from the RLL. Tracing back the paths of these two nodes to the carina identifies the set of branches belonging to the RML. Finally the remaining unclassified airways are assigned to the RLL.

We suggest a quality control check where the reclassification of airways to specific lobes may be required because of anatomical variations of the airways tree. For example the presence of a tracheal bronchus, the most common airway tree variant with a prevalence of 1 percent in the general population \cite{dave_prevalence_2014} would result in a simple manual reclassification of the right upper lobe bronchi. Since only the final analysis is dependent on the lobe classification, reclassification can be done at the end without consequence to the pipeline. We implemented high-level functions that allows easy reclassification by indicating the most central airway segment or node for a given lobe.

\subsection{Metrics}

The two key airway-based metrics that have been computed from the derived data using the analytic pipeline are the airway intertapering gradient and airway tortuosity. Intertapering (equation \ref{eqn:intertapering}) describes the difference in average diameter, $\Bar{d}$ to the average of the parent airway $\Bar{d_{p}}$ compared to the parent airway. 
\begin{equation} \label{eqn:intertapering}
    intertapering = \frac{\Bar{d_{p}}-\Bar{d}}{\Bar{d_{p}}}\\
\end{equation}

Where average diameter $\Bar{d_{p}}$ is derived from the consecutive diameter measurements from the airway segment described in section \ref{sec:awy_measure}.

\vspace{1em}

\begin{figurehere}
    \centering
    \includegraphics[width=\columnwidth]{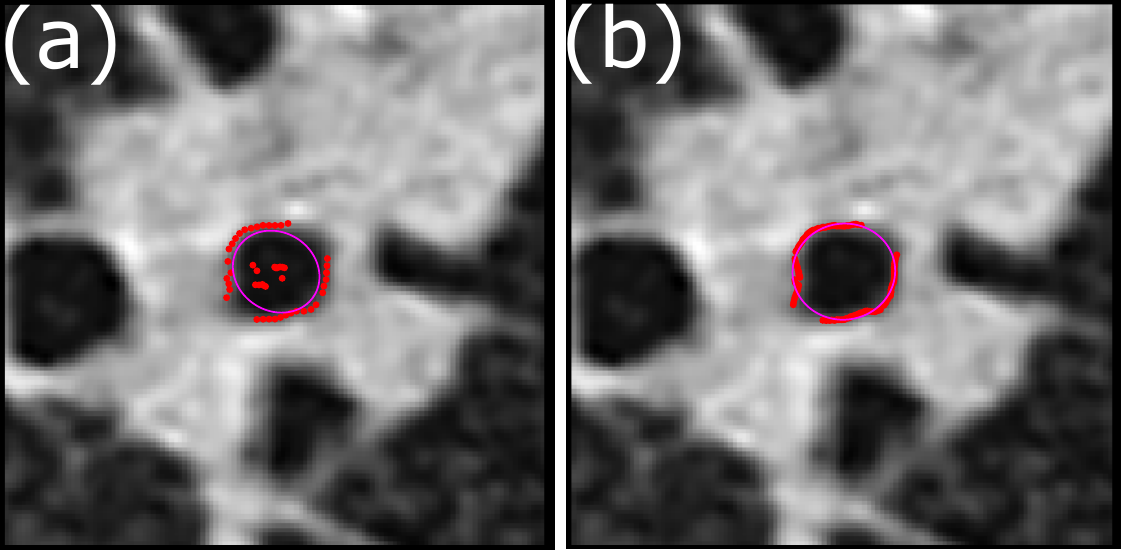}
    \caption{Airway measurement slices of the same perpendicular airway slice. Dots represent inner airway boundary detection of each raycast point. The ellipse is fitted to these points to derive final diameter measurements.
    (a) original method by \cite{quan_tapering_2018} and (b) with our increased number of ray points and addition of outlier removal mechanism. Note that in (a) the ellipse is poorly fitted to the boundary due to raycast points that prematurely stop due to image noise. In (b) These false boundary points are classified as outliers and are therefore removed before fitting the ellipse.}
    \label{fig:outlier}
\end{figurehere}

\begin{algorithm}[H]
\caption{Lobe classification for airways adapted from \cite{gu_automated_2012}. For upper, middle and lower left lobe and right upper lobe.}\label{alg:lobe1}
\begin{algorithmic}
\State Given directed graph of airways, $G(N,A)$ with $N$ nodes and $A$ edges, $n \in N$ and $e \in A$. Arranged in tree like structure from the carina node $n_{C}$. $N$ has positional information associated (x,y,z) in ascending order towards (left, posterior, superior) orientation respectively.
\State $[n_{1}, n_{2}] = E_{o}(n_{C})$ \Comment{out-edges of carina node}
\If{$n_{1}(x) > n_{2}(x)$}
    \State $n_{L} \gets n_{1}$ \Comment{right lung node}
    \State $n_{R} \gets n_{2}$ \Comment{left lung node}
\ElsIf{$n_{2}(x) > n_{1}(x)$}
    \State $n_{L} \gets n_{2}$
    \State $n_{R} \gets n_{1}$
\EndIf
\State $[n_{3}, n_{4}] = E_{o}(n_{L})$
\If{$n_{3}(z) > n_{4}(z)$}
    \State $n_{LUA} \gets n_{3}$ \Comment{Left upper area node}
    \State $n_{LLL} \gets n_{4}$ \Comment{Left lower lobe node}
\ElsIf{$n_{4}(z) > n_{3}(z)$}
    \State $n_{UAL} \gets n_{4}$
    \State $n_{LLL} \gets n_{3}$
\EndIf
\State $[n_{5}, n_{6}] = E_{o}(n_{UAL})$
\If{$n_{5}(z) > n_{6}(z)$}
    \State $n_{LUL} \gets n_{5}$ \Comment{Left upper lobe node}
    \State $n_{LML} \gets n_{6}$ \Comment{Left middle lobe node}
\ElsIf{$n_{4}(z) > n_{3}(z)$}
    \State $n_{LUL} \gets n_{6}$
    \State $n_{LML} \gets n_{5}$
\EndIf\\
\Comment{Left lung now fully classified into lobes}
\State $[n_{7}, n_{8}] = E_{o}(n_{R})$
\If{$n_{7}(z) > n_{8}(z)$}
    \State $n_{RUL} \gets n_{7}$ \Comment{Right upper lobe node}
\ElsIf{$n_{4}(z) > n_{3}(z)$}
    \State $n_{RUL} \gets n_{8}$
\EndIf
\end{algorithmic}
\end{algorithm}

 \begin{algorithm}[H]
\caption{Lobe classification for airways from \cite{gu_automated_2012} for the right middle and lower lobes.}\label{alg:lobe2}
\begin{algorithmic}
 \State Given that upper, middle and lower left lobe and right upper lobe nodes have been classified make subgraph of remaining nodes, $G^{s}(N^{s},A^{s})$ with $N^{s}$ nodes and $A^{s}$ edges, $n^{s} \in N_{s}$ and $e^{s} \in A_{s}$. Carina node, $n_{C}$ is known.
 \State $[N^{s}_{ep}] = E_{endpoints}(G^{s})$ \Comment{Get endpoint nodes}
\State $ V_{z-y}= [N^{s}_{EP}(z)] - [N^{s}_{EP}(y)]$
\State $n^{s}_{RMLep} \gets max(V_{z-y})$ \Comment{Endpoint nodes in right middle lobe}
\State $n^{s}_{RLLep} \gets min(V_{z-y})$ \Comment{Endpoint nodes in right lower lobe}
\State $p_{RML} \gets p(G, n_{C}, n_{RMLep})$ \Comment{p to denote shortest path order in $G$.}
\State $p_{RLL} \gets p(G, n_{C}, n_{RLLep})$
\State $V_{I} \gets intersect((p^{f}_{RML}),(p^{f}_{RLL}))$ \Comment{f to denote path node order reversed}
\State $n_{RML} \gets p^{f}_{RML}[min(V_{I})-1]$
\State RLL assigned to remaining unclassified nodes.
\end{algorithmic}
\end{algorithm}

\begin{figurehere}
    \centering
    \includegraphics[width=\columnwidth]{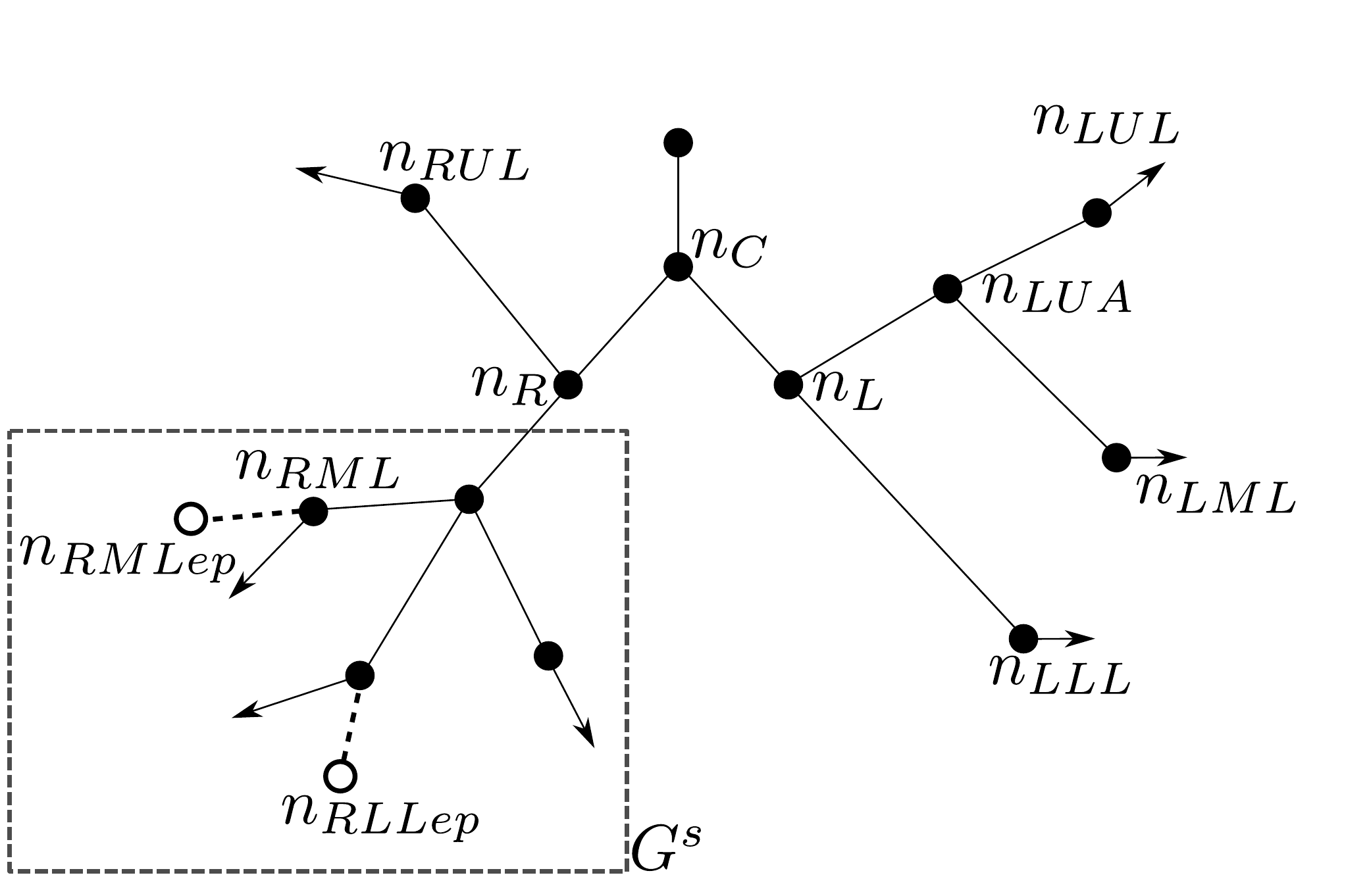}
    \caption{Schematic showing typical airway lobe structure used to automatically classify airways into their lobes. Key airway-tree nodes that are identified in the automated lobe classification algorithm are labelled. The airways arising after a given lobe node are classified into that lobe. Hollow nodes are end-points at the extreme of the airway-tree. Solid lines indicate two nodes are connected by one airway segment. Dashed lines indicate multiple airway segments between two given nodes. Arrows indicate airways that extend beyond the boundaries of the schematic. $G^{s}$ represents the airways considered in Algorithm \ref{alg:lobe2} to identify the right middle and right lower lobes. $n_{C}$, carina node; $n_{R}$, right lung node; $n_{RML}$, right middle lobe node; $n_{RMLep}$, right middle lobe end point; $n_{RLLep}$, right lower lobe end point; $n_{L}$, left lung node; $n_{LUA}$, left upper area node; $n_{LUL}$, left upper lobe node; $n_{LML}$, left middle lobe node; $n_{LLL}$, left lower lobe node.}
    \label{fig:lobeclassgraph}
\end{figurehere}

\vspace{2em}

Airway tortuosity (equation \ref{eqn:tortuosity}) describes the arc-length of an airway segment, $L_{a}$ expressed as a ratio of its euclidean segmental length, $L_{e}$.

\begin{equation}    \label{eqn:tortuosity}
    tortuosity = \frac{L_{a}}{L_{e}}
\end{equation}

Where euclidean segmental length $L_{e}$ is derived from taking the euclidean (straight line) distance between the start and end nodes of the airway segment. Both measures are dimensionless.

\subsubsection{Normal generational range}
The first study analysis aimed to distinguish the assessment of disease severity from disease extent. This is done by truncating airway data to the middle 50\% of airway generations discovered across all healthy volunteers for each lobe. Henceforth, this range of airway generations from the 25th to 75th percentile is referred to as the 'normal generation range', exact values shown in table \ref{tab:normalrange}.

The airways conforming to the normal generation range were evaluated in IPF patients and healthy volunteers in two ways: namely at the airway level and then the patient level. For the airway level analysis, all individual lobar airways within the normal generation range were compared between IPF patients and healthy volunteers for both the airway intertapering gradient (Table \ref{tab:inter_nrm_all}) and airway tortuosity (Table \ref{tab:tor_nrm_all}). For the patient level analysis, the median lobar intertapering gradient (Table \ref{tab:inter_nrm_median}) and airway tortuosity values (Table \ref{tab:tor_nrm_median}) were calculated at an individual patient level and compared between IPF patients and healthy volunteers.

\subsubsection{Central and peripheral ranges}
Two final analyses compared airways at the patient level, but categorised according to the generational level of the airways. The first analysis compared the median lobar airway intertapering gradient (Table \ref{tab:inter_2_6_median}) and airway tortuosity (Table \ref{tab:tor_2_6_median}) for airway generations 2-6. Airways in generation 2-6 can be segmented by most lung airway segmentation algorithms making the results of our airway analyses relatively independent of the quality of the segmentation tool. The second analysis eschewed the normal generation range and compared at a patient level the median lobar airway intertapering gradient (Table \ref{tab:inter_7_p_median}) and airway tortuosity values (Table \ref{tab:tor_7_p_median}) for airway generations 7 and beyond.

\subsection{Deep learning Regression}
We also considered whether an end-to-end deep learning model could learn and therefore predict the median intertapering and tortuosity value of each lobe for the 2nd - 6th generations in IPF patients.

We ran AirQuant on 183 IPF cases from the same hospitals as the main data in section \ref{sec:data} to generate a training dataset. We considered reference-quality deep learning models Alexnet \cite{krizhevsky_imagenet_2012} and VGG16 \cite{simonyan_very_2015} for our network architecture, adapting the final layer to a linear function to suit the regression task. Implemented in PyTorch \cite{paszke_pytorch_2019} with mean squared error loss and AdamW \cite{loshchilov_decoupled_2019} optimiser. We fed in 3D CT images of the lungs for the model to regress to six lobar measures of airway intertapering and six lobar measures of airway tortuosity.

Each image was preprocessed by cropping to a lung mask generated using the trained model by \cite{hofmanninger_automatic_2020} and cubic resampling of all images to achieve a voxel size of $1x1x1 mm$. A centre crop is then applied to achieve a $256x256x256$ 3D input block. CT Hounsfield units are standardised such that the upper and lower limits of a lung window at 250 and -1250 are scaled to 1 and -1. Each regression metric, $x$ is also z-standardised by the training sample's mean and standard deviation. These standardisation strategies as well as the implementation of batch normalisation \cite{ioffe_batch_2015} were implemented to support faster and more stable training and model convergence.

\begin{tablehere}
\caption{Lower and upper quartile of lobar generation occurrence across healthy participants used to inform the 'normal generation range'.}
\centering
\begin{tabular}{p{2.5cm} | p{2.5cm} p{2.5cm}} \label{tab:normalrange}
    Lobe & 25th Percentile & 75th Percentile \\ \hline\hline
    Right upper & 4 & 6 \\
    Left upper & 4 & 5 \\
    Right middle & 3 & 5 \\
    Left middle & 3 & 5 \\
    Right lower & 4 & 7 \\
    Left lower & 4 & 6 \\
    \bottomrule
\end{tabular}
\end{tablehere}
\vspace{2em}

Data augmentation was applied on the CT input with random Gaussian noise and random affine transformation with the exception of scaling. Methods were implemented using the torchio library for medical image deep learning \cite{perez-garcia_torchio_2021}.

We trained using five-fold cross validation and empirically set hyperparameters within limits, learning rate $0.01-0.00001$, batchsize $2-8$, weight decay regularisation $0.0001-0.5$ and dropout after dense layers of $0.1$. 

All training was implemented on 32 GB NVIDIA Tesla V100-DGXS cards.

\subsection{Statistical Analysis}

Airway values for healthy volunteers and IPF patients are compared using the non-parametric Mann-Whitney U test. The median values for the various airway metrics are compared across equivalent lobes and airway generational ranges between the two study groups. 
As this is a pilot study of 28 individuals, multiple comparison corrections are not considered in the main text. These results, which do not alter any conclusions in the paper are shown in the supplementary material.

\section{Results} 
\begin{table*}[!t]
\caption{\label{tab:inter_nrm_all}Intertapering of every investigated airway within the normal generation range}

\begin{tabularx}{\textwidth}{lXcccXcccXd{2.5}}
\toprule
\multicolumn{2}{c}{ } & \multicolumn{3}{c}{Normal} & \multicolumn{1}{c}{ } & \multicolumn{3}{c}{Idiopathic pulmonary fibrosis} & \multicolumn{2}{c}{ } \\
\cmidrule(l{3pt}r{3pt}){3-5} \cmidrule(l{3pt}r{3pt}){7-9}
Lobe & & n & Median & Interquartile range & & n & Median &  Interquartile range & & \multicolumn{1}{c}{p-value}\\
\midrule
Right upper & & 323 & 35.66 & 15.09 &  & 336 & 32.53 & 16.59 &  & 0.00007 \\
Left upper  & & 165 & 36.89 & 21.20 &  & 160 & 32.76 & 18.19 &  & 0.00372 \\
Right middle &  & 111 & 37.28 & 16.84 &  & 155 & 28.95 & 19.07 &  & < 0.00001 \\
Left middle & & 84 & 39.75 & 17.13 &  & 99 & 32.23 & 15.94 &  & 0.00039 \\
Right lower & & 374 & 37.09 & 21.09 &  & 412 & 30.00 & 20.57 &  & < 0.00001 \\
Left lower & & 308 & 40.38 & 20.26 &  & 276 & 31.81 & 21.19 &  & < 0.00001 \\
\bottomrule
\end{tabularx}

\end{table*}

\begin{table*}[!t]
\caption{\label{tab:tor_nrm_all}Tortuosity of every investigated airway within the normal generation range}
\begin{tabularx}{\textwidth}{lXcccXcccXd{2.5}}
\toprule
\multicolumn{2}{c}{ } & \multicolumn{3}{c}{Normal} & \multicolumn{1}{c}{ } & \multicolumn{3}{c}{Idiopathic pulmonary fibrosis} & \multicolumn{2}{c}{ } \\
\cmidrule(l{3pt}r{3pt}){3-5} \cmidrule(l{3pt}r{3pt}){7-9}
Lobe  & & n & Median & Interquartile range & &  n & Median & Interquartile range & & \multicolumn{1}{c}{p-value}\\
\midrule
Right upper  & & 323 & 1.03 & 0.02  & & 336 & 1.03 & 0.03 & & 0.00124 \\
Left upper  & & 165 & 1.02 & 0.02 & &  160 & 1.03 & 0.03 & & 0.00131 \\
Right middle  & & 111 & 1.03 & 0.02 & &  155 & 1.03 & 0.02  & & 0.03380 \\
Left middle  & & 84 & 1.03 & 0.02 & & 99 & 1.03 & 0.03 & & 0.02630 \\
Right lower  & & 374 & 1.03 & 0.02 & &  412 & 1.03 & 0.03 & & < 0.00001 \\
Left lower  & & 308 & 1.03 & 0.02 & &  276 & 1.03 & 0.03 & & < 0.00001 \\
\bottomrule
\end{tabularx}
\end{table*}

\begin{table*}[!t]
\caption{\label{tab:inter_nrm_median}Median lobar intertapering per case within the normal generation range}
\begin{tabularx}{\textwidth}{lXcccXcccXd{2.5}}
\toprule
\multicolumn{2}{c}{ } & \multicolumn{3}{c}{Normal} & \multicolumn{1}{c}{ } & \multicolumn{3}{c}{Idiopathic pulmonary fibrosis} & \multicolumn{2}{c}{ }
\\
\cmidrule(l{3pt}r{3pt}){3-5} \cmidrule(l{3pt}r{3pt}){7-9}
Lobe & & n & Median & Interquartile range & & n & Median & Interquartile range & & \multicolumn{1}{c}{p-value}\\
\midrule
Right upper &  & 14 & 36.58 & 4.34 & & 14  & 32.74 & 7.23 &  & 0.05560\\
Left upper &  & 14 & 37.79 & 7.23 & & 14  & 34.23 & 6.34 &  & 0.16400\\
Right middle &  & 14 & 37.19 & 7.93 &  & 14 & 30.03 & 7.30 &  & 0.00416 \\
Left middle &  & 14 & 40.26 & 5.24 &  & 14 & 33.34 & 8.67 &  & 0.00027 \\
Right lower &  & 14 & 38.17 & 3.97  & & 14 & 29.94 & 4.93 &  & 0.00208 \\
Left lower & & 14 & 39.94 & 5.19 &  & 14 & 32.08 & 8.26 &  & 0.00021 \\
\bottomrule
\end{tabularx}
\end{table*}

\begin{table*}[!t]

\caption{\label{tab:tor_nrm_median}Median lobar tortuosity per case within the normal generation range}
\begin{tabularx}{\textwidth}{lXcccXcccXd{2.5}}
\toprule
\multicolumn{2}{c}{ } & \multicolumn{3}{c}{Normal} & \multicolumn{1}{c}{ } & \multicolumn{3}{c}{Idiopathic pulmonary fibrosis} & \multicolumn{2}{c}{ }
 \\
\cmidrule(l{3pt}r{3pt}){3-5} \cmidrule(l{3pt}r{3pt}){7-9}
Lobe & & n & Median & Interquartile range & & n & Median & Interquartile range & & \multicolumn{1}{c}{p-value} \\
\midrule
Right upper &  & 14 & 1.03 & 0.01 &  & 14 & 1.03 & 0.01 &  & 0.19400 \\
Left upper  & & 14 & 1.02 & 0.01 &  & 14 & 1.03 & 0.01 & &  0.00173 \\
Right middle  & & 14 & 1.03 & 0.01 &  & 14 & 1.03 & 0.01 & &  0.21000\\
Left middle  & & 14 & 1.03 & 0.01 &  & 14 & 1.04 & 0.01 & &  0.03500\\
Right lower  & & 14 & 1.03 & 0.00 &  & 14 & 1.03 & 0.01 & &  0.00173 \\
Left lower  & & 14 & 1.03 & 0.00  & & 14 & 1.03 & 0.00 & &  0.00143 \\
\bottomrule
\end{tabularx}
\end{table*}

Following successful segmentation, AirQuant analyses were fully executed on all airway segments identified.
Failure in the pipeline only occurred at the lobe classification phase. All lobe classifications were visually checked and manually corrected if necessary before further analysis. To summarise, two IPF and two normal cases had LUL airways mislabelled as lingular airways. One IPF case had RLL airways mislabelled as RUL airways. One healthy volunteer exhibited an anatomical variant whereby two RUL branches originated from the right main bronchus and these were mislabelled. Additional lobe results are included in the supplement. 
Each case takes 4-6 hours to process in MATLAB on a 3 GHz Intel-i7 9th gen. processor with 16GB memory workstation. The only manual involvement was of very brief lobe classification corrections to six anomalous cases.

\subsection{Quantitative Airway Analyses}
\begin{figure*}[!t]
    \centering
    \includegraphics[width=\textwidth]{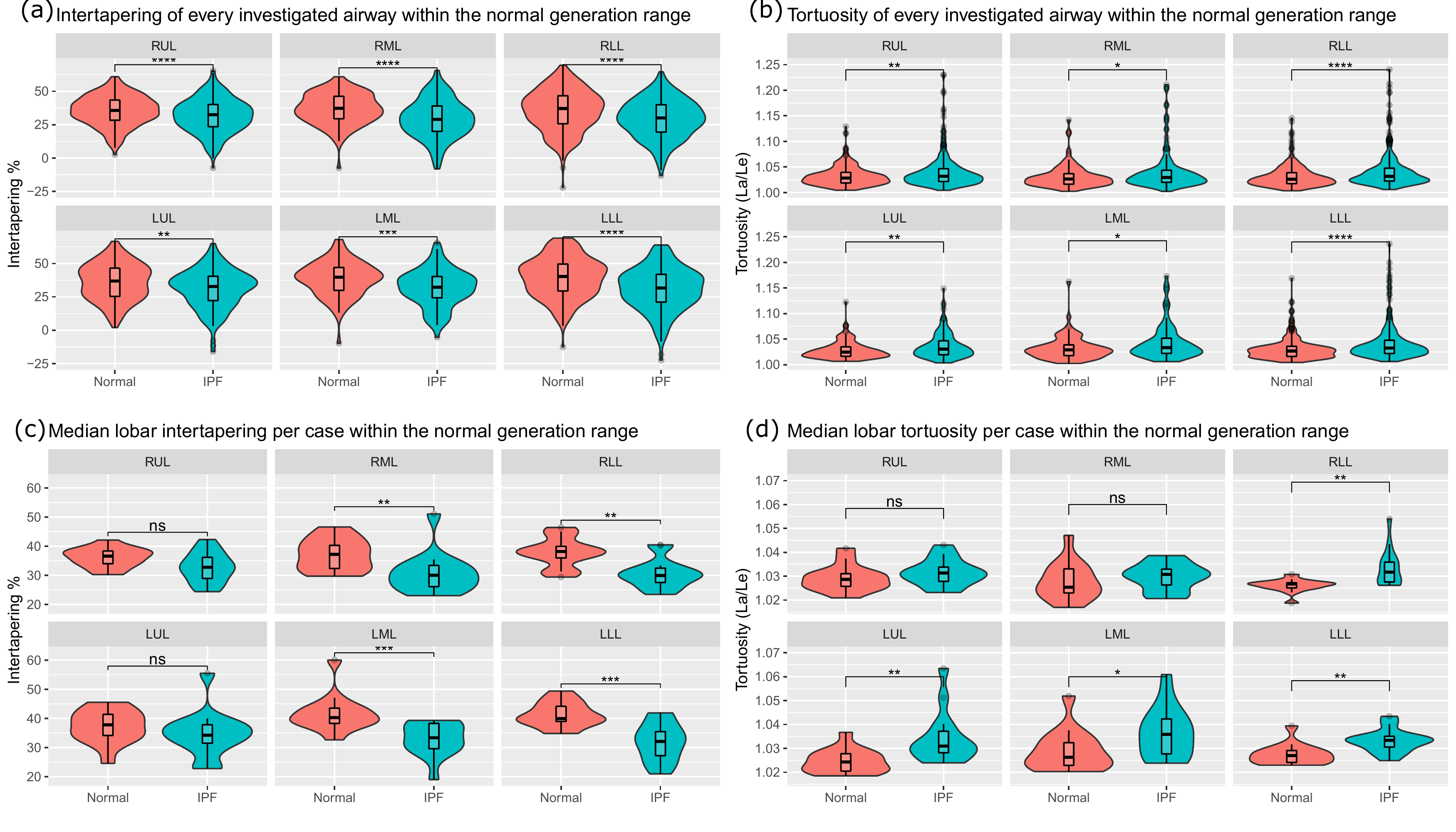}
    \caption{Violin-box plots comparing normal healthy participants and idiopathic pulmonary fibrosis (IPF) patients. The level of significance for the Mann-Whitney U tests for each lobe is shown above the respective plot. (a) and (b) considers every airway segment from every subject in each group. (c) and (d) considers the median lobar values on a per patient basis within the normal generational range. RUL=right upper lobe, LUL=left upper lobe, RML=right middle lobe, LML=left middle lobe, RLL=right lower lobe, LLL=left lower lobe. ns, not significant, *$p<0.05$, **$p<0.01$, ***$p<0.001$, ****$p<0.0001$ in Mann-Whitney U comparison tests.}
    \label{fig:tech_results}
\end{figure*}


When the intertapering gradient was evaluated at the airway level across the normal generation range, a significant reduction ($p\leq0.004$) in intertapering gradient was found in IPF patients compared to healthy volunteers. Airways from patients with IPF were shown to taper less than airways in healthy volunteers (Table \ref{tab:inter_nrm_all}, Figure \ref{fig:tech_results}(a)). Airway tortuosity was significantly greater ($p\leq0.034$) in IPF patients than healthy volunteers (Table \ref{tab:tor_nrm_all}, Figure \ref{fig:tech_results}(b)). Differences were most marked within the lower lobes, in keeping with the typical distribution of IPF airway abnormalities.

For airway changes examined at the patient level, significant reductions in intertapering were seen in IPF patients in the middle ($p\leq0.004$) and lower lobes ($p\leq0.002$) compared to healthy volunteers (Table \ref{tab:inter_nrm_median}, Figure \ref{fig:tech_results}(c)). Airway tortuosity was similarly significantly increased in the LML  ($p\leq0.035$) and both lower lobes  ($p\leq0.002$) (Table \ref{tab:tor_nrm_median}, Figure \ref{fig:tech_results}(d)). The results reflect the lower zone distribution of disease typically seen in IPF patients.

\subsection{Clinical Demonstration on Patients}

\begin{figure*}[!t]
    \centering    \includegraphics[width=\textwidth]{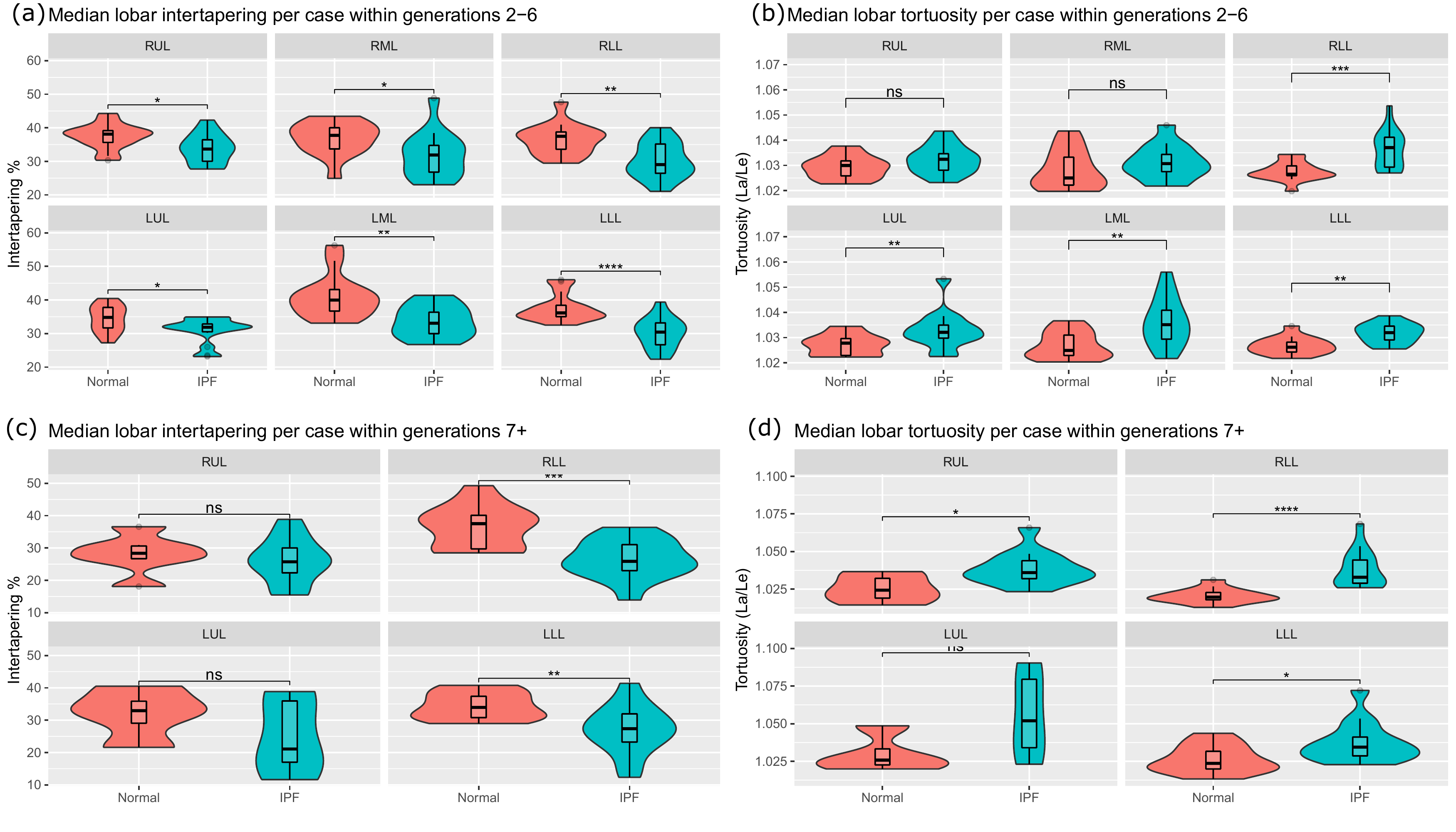}
    \caption{Violin-box plots comparing airways on a per patient basis in normal healthy participants and idiopathic pulmonary fibrosis (IPF) patients. The level of significance of the Mann-Whitney U test for each lobe is shown above the respective plot. (a) and (b) consider airway segments between generations two and six. (c) and (d) only consider airway segments from generation 7 onwards. RUL=right upper lobe, LUL=left upper lobe, RML=right middle lobe, LML=left middle lobe, RLL=right lower lobe, LLL=left lower lobe. ns, not significant, *$p<0.05$, **$p<0.01$, ***$p<0.001$, ****$p<0.0001$ in Mann-Whitney U comparison tests.}
    \label{fig:clin_results}
\end{figure*}

\begin{table*}[!t]
\caption{\label{tab:inter_2_6_median}Median lobar intertapering per case within generations 2-6}
\begin{tabularx}{\textwidth}{lXcccXcccXd{2.5}}
\toprule
\multicolumn{2}{c}{ } & \multicolumn{3}{c}{Normal} & \multicolumn{1}{c}{ } & \multicolumn{3}{c}{Idiopathic pulmonary fibrosis} & \multicolumn{2}{c}{ }
 \\
\cmidrule(l{3pt}r{3pt}){3-5} \cmidrule(l{3pt}r{3pt}){7-9}
Lobe & & n & Median & Interquartile range & & n & Median & Interquartile range & & \multicolumn{1}{c}{p-value} \\

\midrule
Right upper & & 14 & 38.11 & 3.50 & &  14 & 33.66 & 6.41 & &  0.02120 \\
Left upper &  & 14 & 34.80 & 6.08  & & 14 & 31.86 & 2.32 & &  0.04440 \\
Right middle &  & 14 & 37.75 & 6.32 & &  14 & 31.88 & 7.97 & &  0.02410 \\
Left middle &  & 14 & 39.98 & 6.43 & &  14 & 33.07 & 6.38 & &  0.00208 \\
Right lower &  & 14 & 37.46 & 5.16 & &  14 & 29.03 & 8.71 & &  0.00296 \\
Left lower & &  14  & 36.15 & 3.40 &  & 14 & 30.43 & 6.51 & & 0.00003 \\
\bottomrule
\end{tabularx}
\end{table*}

\begin{table*}[!t]
\caption{\label{tab:tor_2_6_median}Median lobar tortuosity per case within generations 2-6}
\begin{tabularx}{\textwidth}{lXcccXcccXd{2.5}}
\toprule
\multicolumn{2}{c}{ } & \multicolumn{3}{c}{Normal} & \multicolumn{1}{c}{ } & \multicolumn{3}{c}{Idiopathic pulmonary fibrosis} & \multicolumn{2}{c}{ }
 \\
\cmidrule(l{3pt}r{3pt}){3-5} \cmidrule(l{3pt}r{3pt}){7-9}
Lobe & & n & Median & Interquartile range & & n & Median & Interquartile range & & \multicolumn{1}{c}{p-value} \\
\midrule
Right upper & &  14 & 1.03 & 0.01 & &  14 & 1.03 & 0.01 & &  0.13700 \\
Left upper & &  14 & 1.03 & 0.01 & &  14 & 1.03 & 0.01 & &  0.00915 \\
Right middle & &  14 & 1.02 & 0.01 & &  14 & 1.03 & 0.01 & &  0.15000 \\
Left middle & &  14 & 1.02 & 0.01 & &  14 & 1.04 & 0.01 & &  0.00674 \\
Right lower & &  14 & 1.03 & 0.00 & &  14 & 1.04 & 0.01 & &  0.00065 \\
Left lower & &  14 & 1.03 & 0.00 & &  14 & 1.03 & 0.01 & &  0.00208 \\
\bottomrule
\end{tabularx}
\end{table*}

\begin{table*}[!t]
\caption{\label{tab:inter_7_p_median}Median lobar intertapering per case within generations 7+}
\begin{tabularx}{\textwidth}{lXcccXcccXd{2.5}}
\toprule
\multicolumn{2}{c}{ } & \multicolumn{3}{c}{Normal} & \multicolumn{1}{c}{ } & \multicolumn{3}{c}{Idiopathic pulmonary fibrosis} & \multicolumn{2}{c}{ }
 \\
\cmidrule(l{3pt}r{3pt}){3-5} \cmidrule(l{3pt}r{3pt}){7-9}
Lobe & & n & Median & Interquartile range & & n & Median & Interquartile range & & \multicolumn{1}{c}{p-value} \\
\midrule
Right upper & &  6 & 28.36 & 3.91 &  & 12 & 25.72 & 7.71 & &  0.43700 \\
Left upper &  & 4 & 32.93 & 6.79 &  & 5 & 21.09 & 18.95 &  & 0.41300 \\
Right lower & &  14 & 37.50 & 10.39 &  & 14 & 25.87 & 8.04 & &  0.00042 \\
Left lower &  & 12 & 33.95 & 6.61 &  & 14 & 27.36 & 8.72 &  & 0.00448 \\
\bottomrule
\end{tabularx}
\end{table*}

\begin{table*}[!t]
\caption{\label{tab:tor_7_p_median}Median lobar tortuosity per case within generations 7+}
\begin{tabularx}{\textwidth}{lXcccXcccXd{2.5}}
\toprule
\multicolumn{2}{c}{ } & \multicolumn{3}{c}{Normal} & \multicolumn{1}{c}{ } & \multicolumn{3}{c}{Idiopathic pulmonary fibrosis} & \multicolumn{2}{c}{ }
 \\
\cmidrule(l{3pt}r{3pt}){3-5} \cmidrule(l{3pt}r{3pt}){7-9}
Lobe & & n & Median & Interquartile range & & n & Median & Interquartile range & & \multicolumn{1}{c}{p-value} \\
\midrule
Right upper  & & 6 & 1.02 & 0.01 &  & 12 & 1.04 & 0.01 & &  0.02450 \\
Left upper &  & 4 & 1.03 & 0.01 &  & 5 & 1.05 & 0.05 &  & 0.19000 \\
Right lower &  & 14 & 1.02 & 0.00 &  & 14 & 1.03 & 0.02 & &  < 0.00001 \\
Left lower &  & 12 & 1.02 & 0.01 &  & 14 & 1.03 & 0.01 &  & 0.02340\\
\bottomrule
\end{tabularx}
\end{table*}

For airway metrics examined at a patient level for generations 2-6, the segmental intertapering value was significantly reduced ($p\leq0.044$) in IPF patients in all lobes. Differences were most marked within the lower lobes  ($p\leq0.003$) (Table \ref{tab:inter_2_6_median}, Figure \ref{fig:clin_results}(a)). Segmental airway tortuosity was significantly increased in left-sided lung lobes ($p\leq0.009$) and the RLL  ($p\leq0.001$)(Table \ref{tab:tor_2_6_median}, Figure \ref{fig:clin_results}(b)).

Airways at generations 7 and beyond were not routinely seen in the middle lobes. Accordingly airway metrics were only examined in the upper and lower lobes. Significant differences in segmental intertapering  ($p\leq0.004$) (Table \ref{tab:inter_7_p_median}, Figure \ref{fig:clin_results}(c)) and segmental airway tortuosity  ($p\leq0.023$)(Table \ref{tab:tor_7_p_median}, Figure \ref{fig:clin_results}(d)) were seen in the lower lobes between IPF patients and healthy volunteers.


\subsection{Deep learning Regression}
\begin{figure*}[!t]
    \centering
    \includegraphics[width=\textwidth]{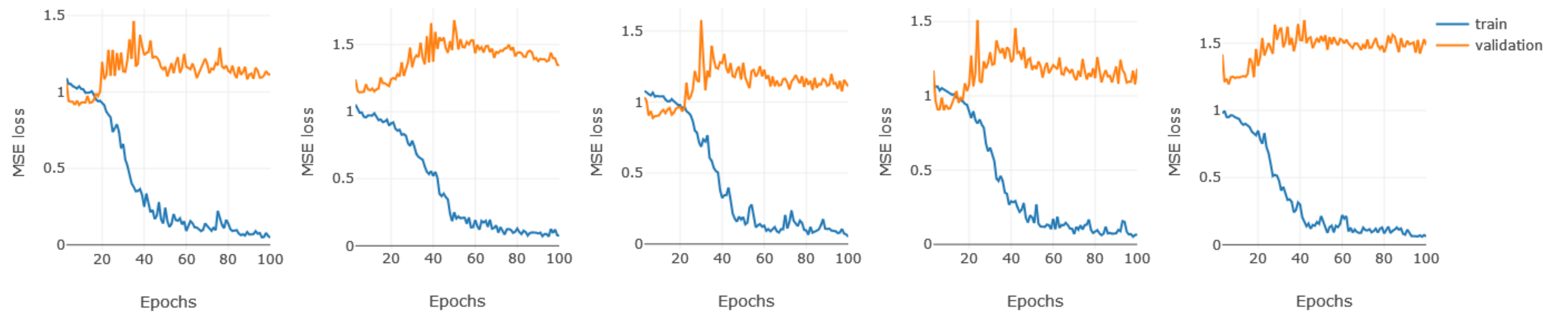}
    \caption{Mean squared error (MSE) training and validation loss curves for the deep learning regression task to AirQuant computed variables. Training curves are shown for each data fold in cross validation training of the VGG16 model, demonstrating that though the models converge in training they do not generalise in validation.}
    \label{fig:losscurve}
\end{figure*}

Five fold training and validation curves shown in Figure \ref{fig:losscurve} for an example case with VGG16 model hyperparameters set to learning rate $0.0001$, batch size 8, dropout 0.1 and weight decay 0.5.
The average R-squared values across the considered 12 variables, averaged across the 5 fold validation sets was $(0.00\pm0.07)$.
Though the training loss was found to converge, this did not occur in the validation dataset with varying degrees of loss regularisation. Therefore the model was deemed unable to generalise beyond the training dataset. 

\section{Discussion}
Our results highlight the potential for CT-based morphological airway measurements to identify differences in disease extent and severity of traction bronchiectasis in patients with IPF. To date airway damage in IPF has been thought to be a primarily distal airway phenomenon. Our results have supported this by showing that the largest distinction in intertapering and tortuosity between healthy volunteers and IPF patients was seen in the most distal airway generations. Yet the identification of a reduction in airway intertapering and increase in tortuosity in proximal airway branches suggests that AirQuant could provide valuable insights when applied across a wide range of the airway tree, meaning that it can be usefully applied to most airway segmentation methods.


Our findings are comparable to previous analyses of intertapering gradients of proximal airways in patients with idiopathic bronchiectasis and healthy controls \cite{kuo_airway_2020}. Our results diverged when more distal airways were evaluated, reflecting the ability of our 2D-UNET segmentation to identify larger numbers of more distal airway branches. The primary advantage of our method lies in its automation which allows hundreds of airways in a single patient to be analysed, and therefore makes evaluation of large patient cohorts feasible. 

The concept that traction bronchiectasis scored visually on CT imaging could be a prognostic variable in IPF patients was first described almost 15 years ago \cite{sumikawa_computed_2008, edey_fibrotic_2011}. More recently, change in visual traction bronchiectasis scores has been shown to identify disease progression on longitudinal IPF imaging \cite{jacob_serial_2020}. The challenges associated with employing visual traction bronchiectasis estimation however are many. These include the time consuming nature of visual analysis, the requirement for expert reads of the images, where experts are typically in short supply, are expensive to employ and are prone to interobserver variation \cite{jacob_automated_2016}.  The potential that automating airway tapering assessments might alleviate some of the challenges associated with visual CT estimation of traction bronchiectasis by providing an objective, rapid and sensitive measure of lung disease severity in IPF was the motivation behind the current study. 

AirQuant is reliant on an airway lumen segmentation. 
Our pipeline can be implemented following any airway segmentation method. We chose to use automated segmentation methods, specifically to demonstrate that meaningful measures can be derived using AirQuant without recourse to labour intensive manual labelling of the extensive airway network.
Our results are influenced by the airway segmentation algorithm and cannot be isolated from the underlying segmentation. Yet the statistically significant results shown for airway generations 2-6 indicates that disease severity metrics can be assessed using most airway segmentation algorithms.

The failure of the deep learning regression models to generalise to these variables derived from AirQuant suggests that a straight-forward deep learning approach is not suitable for predicting these variables in limited data. Furthermore, a naive 3D CT input network would only produce those final median values, losing the detailed information of the airway bronchial tree. Based on these results, we argue for the validity of our end-to-end proposed AirQuant pipeline.

Our deep learning regression approach to predict airway inter-tapering and tortuosity was not able to generalise. This may be a consequence of the sheer complexity of the diverse data being analysed. A lobe as described in our experiments comprises 2nd-6th segmental generation airways. In total, approximately 31 airway segments are contained within this generational range (assuming there are only airway bifurcations rather than trifurcations). The 31 airway segments could comprise a variety of morphologies which may yet produce the same median morphological measure. Accordingly it would be extremely difficult for a model to train without additional contextual information being provided to the network, for example, registration of all images to a common space or paired in a multi-task approach with airway segmentation. 

A further constraint lies in the heterogeneity of the input data which originated from two different centres where the CT imaging was produced on different scanners and reconstructed using a range of kernels. Table \ref{tab:data} highlights the various reconstruction kernels that were used in the training/validation data. Our findings highlight the challenges in applying a deep learning framework to airway morphological analysis and emphasise the utility of the AirQuant pipeline.

Our airway lobe classification algorithm can be affected by anatomical variants although these are relatively uncommon \cite{gu_automated_2012}. In mitigation of this, the lobe classification of individual airways can be easily retrospectively corrected in our pipeline. As future work we aim to improve the robustness with which our lobar classification system deals with anatomical variants which will likely be present in very large data sets ($>>1000$).\\

In the latter airway generations, an AirQuant segmental branch may in fact consist of multiple anatomical segments. This may occur if sibling branches are not identified in the segmentation and therefore not acknowledged as bifurcations. Though unlikely in the central airways, it can affect metrics in higher generations. One idea to tackle this is by a manual review of segments over a particular length, with methods to manually add in nodes to divide up segments. Though this could be a labourious task that would need a well developed suite of tools to facilitate.  \\

Despite having a relatively small sample size of 14 IPF cases we have demonstrated in a proof of concept study that our measurements can distinguish patients with traction bronchiectasis from those without. It will be important to validate the clinical impact of our measurements on larger IPF datasets. 
AirQuant was successful in examining every case, but relies on a good lumen segmentation.
Efforts have been made to make the centreline extraction and graph conversion stage robust to loops. However when coming across branches that cause loops, it is difficult to determine which graph edge is the valid airway branch and which is the anomaly. It is likely that the wrong offending graph edge may have been removed in some cases, as it is not trivial to identify the anomaly in an automated way.


In conclusion, we have demonstrated that the airway intertapering gradient is reduced and airway tortuosity  enhanced in IPF patients compared to healthy participants. The findings were accentuated in the lower lobes which is consistent with the typical distribution of traction bronchiectasis in IPF. Our pilot analyses suggest that automated airway analyses show great promise for the assessment of disease severity and extent in IPF trials and clinical care.

\section*{Acknowledgements}
This study was supported by i4health EPSRC (EP/S021930/1), Cystic Fibrosis Trust (VIA059), Wellcome Trust (209553/Z/17/Z), CMIC Platform EPSRC (EP/M020533/1) and NIHR-UCLH BRC.

\bibliographystyle{ieeetr}
\bibliography{AirQuantPaper}

\end{multicols}
\end{document}